\documentstyle [12pt] {article}
\title{CAN EQUIVALENCE PRINCIPLE BE CONSISTENT WITH THE
  BOHR-SOMERFELD-HANSSON THEORY OF THE NEWTONIAN GRAVITY}

\author{Vladan Pankovi\'c, Darko V. Kapor\\
Department of Physics, Faculty of Sciences, 21000 Novi Sad,\\ Trg
Dositeja Obradovi\'ca 4, Serbia, \\vladan.pankovic@df.uns.ac.rs}

\date {}
\begin {document}
\maketitle \vspace {0.5cm}
\vspace {0.5cm}

\begin {abstract}
In this work we consider some consequences of the
Bohr-Sommerfeld-Hansson (Old or quasi-classical) quantum theory of
the Newtonian gravity, i.e. of the "gravitational atom". We prove
that in this case (for gravitational central force and quantized
angular momentum) centrifugal acceleration becomes
formally-theoretically dependent (proportional to fourth degree)
of the mass of "gravitational electron" rotating around
"gravitational nucleus" for any quantum number (state). It
seemingly leads toward a paradoxical breaking of the relativistic
equivalence principle which contradicts to real experimental data.
We demonstrate that this equivalence principle breaking does not
really appear in the (quasi classical) quantum theory, but that it
necessary appears only in a hypothetical extension of the quantum
theory that needs a classical like interpretation of the
Bohr-Sommerfeld angular momentum quantization postulate. It is, in
some sense, similar to Bell-Aspect analysis that points out that a
hypothetical deterministic extension of the quantum mechanics, in
distinction to usual quantum mechanics, can reproduce experimental
data if and only if it is non-local (superluminal) in
contradiction with relativistic locality (luminality) principle.
\end {abstract}

Equivalence principle, or simply equivalence between
gravitational, i.e. heavy and inertial mass, represents, as it is
well-known [1], [2], the basic principle (cornerstone) of the
general theory of relativity. But, also, this principle appears
phenomenologically already in the Newtonian classical mechanics
and gravitation theory. Incorporation of the equivalence principle
in the quantum field theory or development of the quantum theory
of gravity represents an extremely hard problem [3] (which,
probably, can be solved only within string theories). However, it
is shown experimentally [4] that equivalence principle can be
successfully applied at least in the domains of the
non-relativistic quantum mechanics and low energetic sector of the
quantum field theory. It implies that equivalence principle can be
satisfactorily used in such domains where quasi-classical
approximation of the quantum mechanics or Old, Bohr-Sommerfeld
quantum theory of the atom can be applied.

In this work we shall consider some consequences of the
Bohr-Sommerfeld-Hansson (Old or quasi-classical) quantum theory of
the Newtonian gravity, i.e. of the "gravitational atom"
(super-system consisting of the heavy, practically unmovable,
central system, "gravitational nucleus", classically
gravitationally interacting with small, periferical probe system,
"gravitational electron", rotating stablely around "gravitational
nucleus") [5]. We prove that in this case (for gravitational
central force and quantized angular momentum) centrifugal
acceleration becomes formally-theoretically dependent
(proportional to fourth degree) of the mass of "gravitational
electron" for any quantum number (state). It seemingly leads
toward a paradoxical breaking of the relativistic equivalence
principle which contradicts to mentioned, real experimental data.
We demonstrate that this equivalence principle breaking does not
really appear in the (quasi classical) quantum theory, but that it
necessary appears only in a hypothetical extension of the quantum
theory that needs a classical like interpretation of the
Bohr-Sommerfeld angular momentum quantization postulate. It is, in
some sense, similar to Bell-Aspect analysis [6], [7], that points
out that a hypothetical deterministic extension of the quantum
mechanics, in distinction to usual quantum mechanics, can
reproduce experimental data if and only if it is non-local
(superluminal) in contradiction with relativistic locality
(luminality) principle.

Suppose that there is an attractive classical physical central
Kepler force
\begin {equation}
  F = \frac {A}{R^{2}}
\end {equation}
where A represents corresponding parameter and R distance in
respect to source of the force in the coordinate beginning.

Suppose that this force acts on a small, periferical probe system
with mass m so that this system rotates stablely around coordinate
beginning with a constant speed v. In this case there is classical
mechanical equivalence between given attractive force (1) and
centrifugal force
\begin {equation}
   \frac {mv^{2}}{R} = \frac {A}{R^{2}}
\end {equation}
that implies the following centrifugal acceleration
\begin {equation}
   {\it a} \equiv \frac {v^{2}}{R} = \frac {A}{mR^{2}}               .
\end {equation}

Suppose, however, that in a more precise quantum theoretical
description, within domains of the Old, Bohr-Sommerfeld quantum
theory, mentioned rotation of small system satisfies additionally
Bohr-Sommerfeld angular momentum quantization postulate
\begin {equation}
   mvR = n\hbar
\end {equation}
where $\hbar$ represents the reduced Planck constant, and n -
quantum number for $n=1,2,...$ .

Equations system (2), (4) has simple solution
\begin {equation}
  R_{n} = \frac {n^{2}\hbar^{2}}{Am}
\end {equation}
\begin {equation}
  v_{n} = \frac {A}{n\hbar}
\end {equation}
for $n=1,2,...$ .

It implies the following centrifugal acceleration
\begin {equation}
   {\it a}_{n}\equiv \frac {v^{2}_{n}}{R_{n}}=  \frac {A^{3}m}{n^{4}\hbar^{4}}
\end {equation}
for $n=1, 2,...$ .

As it is well-known, we paraphrase Feynman [8], concept of the
force and acceleration disappears or, eventually, has only
secondary role within quantum mechanics where energy, momentum and
probability amplitudes or quantum states obtain primary role.
Nevertheless, concept of the centrifugal acceleration, according
to (2)-(7), represents one of the basic approximate concepts of
the Old quantum theory.

It is not hard to see that classical centrifugal acceleration (3)
depends of m in a principally different way in respect to quantum
centrifugal acceleration (7).

Consider a case when A does not depend of m, e.g. in the case of
the Coulomb force
\begin {equation}
   A = \frac {e^{2}}{4\pi \epsilon_{0}}
\end {equation}
where e represents the elementary electric charge and e0 - vacuum
electric permittivity. Then classical centrifugal force (3),
according to (8), equals
\begin {equation}
   {\it a} = \frac {e^{2}}{4\pi \epsilon_{0}}\frac {1}{mR^{2}}
\end {equation}
so that it is inversely proportional to m. Simultaneously, quantum
centrifugal acceleration (7), according to (8), equals
\begin {equation}
   {\it a}_{n} =  (\frac {e^{2}}{4\pi \epsilon_{0}})^{3}\frac { m}{n^{4}\hbar^{4}}
\end {equation}
so that it is directly proportional to m, for $n=1, 2,...$ .

Consider other case when A is proportional to m, e.g. in the case
of the Newtonian gravitational force (so that equations system
(2), (4) represents the Hansson formalism of the Newtonian Quantum
Gravity [5])
\begin {equation}
   A = GmM           .
\end {equation}
Here M, much larger than m, represents the mass of a large,
practically unmovable central system, source of the gravitational
force. Also, G represents the Newtonian gravitational constant.

Now we have the following consequences. Classical centrifugal
force (3), according to (11), equals
\begin {equation}
   {\it a} \equiv \frac {v^{2}}{R} = \frac {GM}{R^{2}}
\end {equation}
so that it independent of m in full agreement with equivalence
principle.

Simultaneously, quantum centrifugal acceleration (7), according to
(11), equals
\begin {equation}
   {\it a}_{n}\equiv \frac {v^{2}_{n}}{R_{n}}=  \frac {GM}{R^{2}_{n}} = \frac {G^{3}m^{4}M^{3}}{n^{4}\hbar^{4}}
\end {equation}
so that it strongly depends of m (it is proportional to fourth
degree of m), for $n=1, 2,...$ . At the first sight it represents
a paradoxical result since it practically implies breaking of the
equivalence principle.

Since experimental verification of the equivalence principle [4]
can be considered unambiguous and since deduction of (13) within
Old quantum theoretical applicability domains is
formally-theoretically correct solution of the paradox needs a
physical reinterpretation of the deduction procedure.

Firstly, it can be observed that classical form of the centrifugal
acceleration (12) and Old quantum theoretical form of the
centrifugal acceleration without explication of the dependence of
m
\begin {equation}
   {\it a}_{n}\equiv \frac {v^{2}_{n}}{R_{n}}=  \frac {GM}{R^{2}_{n}}
\end {equation}
for $n=1, 2,...$ are completely analogous. In other words,
classical form of the centrifugal acceleration (12) is absolutely
independent of m, while in quasi-classical form of the centrifugal
acceleration (14) dependence of m is only implicit, over variables
$v_{n}$ and $R_{n}$ for $n=1, 2,...$ .

We shall suppose (and prove later) that real (effective) physical
sense within Old quantum theory has only expression (14) even if,
formally-mathematically, this expression is equivalent to $\frac
{G^{3}m^{4}M^{3}}{n^{4}\hbar^{4}}$.

Such distinction between real (effective) and formal Old quantum
theoretical concepts can seem very strange but similar situation
exists regularly within quantum mechanics, precisely standard
quantum mechanical formalism [9], [10]. Concretely, as it is
well-known, quantum mechanical average value of an observable
$\hat {A}$ in a superposition quantum state $|S>= \sum_{n}
c_{n}|A_{n}>$, where $c_{n}$ for $n=1, 2, …$ represent
superposition coefficients and $|A_{n}>$ and $A_{n}$ for
$n=1,2,... $ eigen states and eigen values of $\hat {A}$, equals
formally-mathematically exactly
\begin {equation}
     <\hat {A}> = S n |c_{n}|^{2}A_{n}
\end {equation}
even if, before measurement, this average value does not exist
really neither superposition $|S>$ represents a impure or mixed
state. It, according to Bell theoretical [6] and Aspect et al [7]
experimental analyses simply means that quantum mechanics cannot
be exactly presented as any kind of a classical mechanical and
similar deterministic theory in a satisfactory and plausible way,
precisely a way that does not contradict to relativistic locality
(luminality) principle. (Precisely, Bell-Aspect analysis points
out that a hypothetical deterministic extension of the quantum
mechanics, in distinction to usual quantum mechanics, can
reproduce experimental data if and only if it is non-local
(superluminal) in contradiction with relativistic locality
(luminality) principle.) For this reason, application of the
approximate classical mechanical concepts within quantum mechanics
is principally limited (by Heisenberg uncertainty principle or
Bohr complementarity principle) conceptually very similarly to
limitation of the approximate classical mechanical concepts within
theory of relativity as it has been many times pointed out by Bohr
[11], [12]. (Simultaneously, hypothetical extension of the quantum
mechanics by hidden variables is very similar to extension of the
theory of relativity by ether concepts.) Bohr, for example,
stated: "Before concluding I should still like to emphasize the
bearing of the great lesson derived from general relativity theory
upon the question of physical reality in the field of quantum
theory. In fact, notwithstanding all characteristic differences,
the situation we are concerned with in these generalizations of
classical theory presents striking analogies which have often been
noted. Especially, the singular position of measuring instrument
in the account of quantum phenomena, just discussed, appears
closely analogous to the well-known necessity in relativity theory
of upholding an ordinary description of all measuring processes,
including sharp distinction between space and time coordinates,
although very essence of this theory is the establishment of new
physical laws, in comprehension of which we must renounce the
customary separation of space and time ideas." [11]

So, within Old quantum theory, as an approximate form of the
quantum mechanics, distinction between real (effective) and
formal-theoretical concepts, or, between quantum and classical
concepts is necessary too. Concretely, expression (2),
representing a classical mechanical rotation stability condition,
can be interpreted completely classical mechanically. But
Bohr-Sommerfeld angular momentum quantization postulate (4) cannot
be interpreted classical mechanically at all. It, as well as
quantum jumps, represents, we paraphrase Bohr [12], an irrational
feature from classical mechanical view point.

All this implies that Old quantum theoretical form of the
centrifugal acceleration without explication of the dependence of
m ($\hbar$ and $n$ for $n=1, 2,...$), representing the direct
consequence of mentioned classical mechanical rotation stability
condition only (without explicit use of the Bohr-Sommerfeld
angular momentum quantization postulate), can be interpreted
completely classical mechanically, including certain satisfaction
of the equivalence principle, on the one hand. On the other hand,
all this implies that numerical value of the centrifugal
acceleration $ \frac {G^{3}m^{4}M^{3}}{n^{4}\hbar^{4}}$ for $n=1,
2,...$ , that includes Bohr-Sommerfeld angular momentum
quantization postulate, cannot be interpreted classical
mechanically at all. For this reason there is no any certain or
real (effective) correspondence between this numerical value and
equivalence principle. Old quantum theory as well as
(non-relativistic) quantum mechanics does not contradict to
relativistic equivalence principle. Only a classical mechanical or
similar attempt of the complete interpretation of Bohr-Sommerfeld
angular momentum quantization postulate, i.e. generally speaking,
an attempt of the classical like extension of the Old quantum
theory can lead toward breaking of the relativistic equivalence
principle, but it contradicts to experimental data [4]. It is, in
some sense, similar to Bell-Aspect analysis [6], [7] that points
out that a hypothetical deterministic extension of the quantum
mechanics, in distinction to usual quantum mechanics, can
reproduce experimental data if and only if it is non-local
(superluminal) in contradiction with relativistic locality
(luminality) principle.

In this way our supposition that real (effective) physical sense
within Old quantum theory has only expression (14) even if,
formally-mathematically, this expression is equivalent to $ \frac
{G^{3}m^{4}M^{3}}{n^{4}\hbar^{4}}$  is definitely proved within
standard quantum mechanical formalism and its quasi-classical
approximation. Here, like to remarkable gedanken (though)
experiment of the photon that leaves a box hanged on an elastic
spring in the Earth gravitational field discussed by Einstein and
Bohr [12], here is no any dynamical contradiction but there is a
full conceptual agreement between quantum mechanics and general
theory of relativity. For this reason we can only repeat the
following Bohr words: "The dependence of the reference system, in
relativity theory, of all readings of scales and clocks may even
be compared with essentially uncontrollable exchange of the
momentum or energy between the objects of measurement and all
instruments defining the space-time system of the reference, which
in quantum theory confront us with the situation characterized by
the notion of complementarity. In fact this new feature of natural
philosophy means a radical revision of our attitude as regards
physical reality, which may be paralleled with the fundamental
modification of all ideas regarding the absolute character of
physical phenomena, brought about general theory of relativity."

In conclusion we can shortly repeat and point out the following.
In this work we consider some consequences of the
Bohr-Sommerfeld-Hansson (Old or quasi-classical) quantum theory of
the Newtonian gravity, i.e. of the "gravitational atom". We prove
that in this case (for gravitational central force and quantized
angular momentum) centrifugal acceleration becomes
formally-theoretically dependent (proportional to fourth degree)
of the mass of "gravitational electron" rotating around
"gravitational nucleus" for any quantum number (state). It
seemingly leads toward a paradoxical breaking of the relativistic
equivalence principle which contradicts to real experimental data.
We demonstrate that this equivalence principle breaking does not
really appear in the (quasi classical) quantum theory, but that it
necessary appears only in a hypothetical extension of the quantum
theory that needs a classical like interpretation of the
Bohr-Sommerfeld angular momentum quantization postulate. It is, in
some sense, similar to Bell-Aspect analysis that points out that a
hypothetical deterministic extension of the quantum mechanics, in
distinction to usual quantum mechanics, can reproduce experimental
data if and only if it is non-local (superluminal) in
contradiction with relativistic locality (luminality) principle.

\vspace{1cm}

{\large \bf References}

\begin{itemize}

\item [[1]]  S.Weinberg, {\it Gravitation and Cosmology - Principles and Applications of the General Theory of Relativity} (John Wiley, New York, 1972)
\item [[2]] M. P. Haugen, C. Lämmerzahl, {\it Principles of Equivalence: Their Role in Gravitation Physics and Experiments that Test Them} (Springer, New York Berlin, 2001)
\item [[3]] {\it General Relativity}, Eds. S. W. Hawking, W. Israel (Cambridge University Press, Cambridge, London, 1979)
\item [[4]] S. Fray, C. A. Diez, T. W. Hänsch, M. Weitz, Phys.Rev. Lett. {\bf 94} (2004) 240404; physics/0411052
\item [[5]] J. Hansson, {\it Newtonian Quantum Gravity} gr-qc/0612025
\item [[6]] J. S. Bell, Physics {\bf 1} (1964) 195
\item [[7]] A. Aspect, P. Grangier, G. Roger, Phys.Rev.Lett. {\bf 47} (1981) 460
\item [[8]] R. P. Feynman, R. B. Leighton, M. Sands, {\it The Feynman Lectures on Physics, Vol. 6} (Addison-Wesley Inc., Reading, Mass. 1963)
\item [[9]]   J. von Neumann,  {\it Mathematische Grundlagen der Quanten Mechanik} (Spiringer Verlag, Berlin, 1932)
\item [[10]] P. A. M. Dirac, {\it Principles of Quantum Mechanics} (Clarendon Press, Oxford, 1958)
\item [[11]] N. Bohr, Phys.Rev. {\bf 48} (1935) 696
\item [[12]] N. Bohr, {\it Atomic Physics and Human Knowledge} (John Wiley, New York, 1958)

\end {itemize}

\end {document}